\documentclass[%
 showpacs,
 showkeys,
 amsmath,amssymb,
]{revtex4-1}

\usepackage{graphicx}
\usepackage{dcolumn}
\usepackage{bm}

\usepackage{CJK}

\usepackage{color}
\usepackage{amssymb}
\usepackage{amsfonts}

\usepackage[colorlinks,urlcolor=blue,linkcolor=blue,citecolor=blue,anchorcolor=blue]{hyperref}

\begin{document}


\preprint{APS/123-QED}

\title{Generating Lieb and super-honeycomb lattices by employing the fractional Talbot effect}

\author{Hua Zhong$^{1,2,3}$}
\author{Yiqi Zhang$^{1,2,3}$}
\email{zhangyiqi@mail.xjtu.edu.cn}
\author{Milivoj R. Beli\'c$^{4}$}
\author{Yanpeng Zhang$^{2}$}
\affiliation{%
 $^1$Key Laboratory for Physical Electronics and Devices of the Ministry of Education \& Shaanxi Key Lab of Information Photonic Technique,
Xi'an Jiaotong University, Xi'an 710049, China \\
$^2$School of Electronic and Information Engineering, Xi'an Jiaotong University, Xi'an, Shaanxi 710049, China \\
$^3$Guangdong Xi'an Jiaotong University Academy, Foshan 528300, China\\
$^4$science Program, Texas A\&M University at Qatar, P.O. Box 23874 Doha, Qatar
}%

\date{\today}

\begin{abstract}
  \noindent
We demonstrate a novel method for producing optically-induced Lieb and super-honeycomb lattices, by employing the fractional Talbot effect of specific periodic beam structures.
Our numerical and analytical results display the generation of Lieb and super-honeycomb
lattices at fractional Talbot lengths effectively and with high beam quality.
By adjusting the initial phase shifts of the interfering beams,
the incident periodic beam structures, as well as the lattices with broken inversion symmetry, can be constructed {\it in situ}.
This research suggests not only a possible practical utilization of  the Talbot effect in the production of novel optically-induced lattices but also in the studies of related optical topological phenomena.
\end{abstract}

\maketitle

%
\section{Introduction}

The Talbot effect is a self-imaging effect that can be generally interpreted by the Fresnel diffraction theory.
It was first observed by H. F. Talbot \cite{talbot.pm.9.401.1836}
and theoretically explained by L. Rayleigh \cite{rayleigh.pm.11.196.1881} in the 19th century.
{Nonetheless, the Talbot effect still attracts considerable attention of many research groups around the globe \cite{wen.aop.5.83.2013}, for} its potential applications in image preprocessing and synthesis, photolithography,
optical testing, optical metrology, spectrometry, and optical computing.
Today, research efforts that involve the Talbot effect include
atomic optics \cite{wen.apl.98.081108.2011,zhang.ieee.4.2057.2012,zhang.pra.97.013603.2018},
quantum optics \cite{song.prl.107.033902.2011,jin.apl.101.211115.2012},
nonlinear optics \cite{zhang.prl.104.183901.2010,zhang.pre.89.032902.2014,zhang.pre.91.032916.2015},
waveguide arrays \cite{iwanow.prl.95.053902.2005},
photonic lattices \cite{ramezani.prl.109.033902.2012},
Bose-Einstein condensates \cite{deng.prl.83.5407.1999,ryu.prl.96.160403.2006}, and
electronics \cite{salas.prb.93.104305.2016}, to name a few.
It is worth mentioning that the Talbot effect can also be observed by using spherical waves \cite{azana.prl.112.213902.2014} and accelerating beams \cite{lumer.prl.115.013901.2015,zhang.ol.40.5742.2015,zhang.ol.41.3273.2016}.

In optics, periodic beam structures can be prepared by employing the multi-beam interference method;
a thorough review of this method can be found in \cite{burrow.micromachines.2.221.2011}.
Generally, such periodic beam structures may not exhibit the Talbot effect, since they represent discrete nondiffracting beam patterns \cite{boguslawski.pra.84.013832.2011}.
But, if the nondiffracting property of such beams is broken, then they may display the Talbot effect during propagation in free space.
Incidentally, these beams may not only display the Talbot effect, but it has been reported that the beams prepared by the multi-beam interference method
can also be used to investigate topological and optical-quantum analogue phenomena \cite{zhang.lpr.9.331.2015,zhang.prl.117.123601.2016}.
It is well known that the periodic beam structures may exhibit surprising patterns (i.e., the fractional Talbot images) in the realization of the Talbot effect.
In previous investigations, the fractional Talbot images were presented only at typical locations (e.g., half or quarter of the Talbot length).
In this work, we discover pattern revivals at atypical fractional lengths, offering new insights and applications of the Talbot effect.

Hence, in this paper we investigate the formation of Talbot patterns of specific periodic beam structures and report that
Lieb \cite{leykam.pra.86.031805.2012,vicencio.prl.114.245503.2015,mukherjee.prl.114.245504.2015,xia.ol.41.1435.2016,diebel.prl.116.183902.2016}
and super-honeycomb lattice arrays \cite{lan.prb.85.155451.2012,zhong.adp.529.1600258.2017,zhu.oe.26.24307.2018} can be produced accordingly.
Recently, such lattices stirred considerable interest, due to their unique band structure and intriguing topological properties.
As far as we know, such lattices cannot be prepared directly using the multi-beam interference method \cite{burrow.micromachines.2.221.2011,xia.ol.41.1435.2016,lan.prb.85.155451.2012}.
Indeed, spatial modulation \cite{song.nc.6.6272.2015} and femtosecond laser writing technique \cite{rechtsman.nature.496.196.2013} can nowadays produce almost any photonic structure.
However, it would be of interest to find more economic and feasible way,
and this is the motivation for our research.
Accordingly, we find that such lattices can be formed using the Talbot images of certain periodic beams at fractional Talbot lengths.
This discovery not only provides a new avenue for producing various lattices that can be utilized in photonic research,
but also broadens the base of potential applications of the Talbot effect.

\section{Basic model}
The model that describes beam propagation in free space, {can be written as the dimensionless} Schr\"odinger-like paraxial wave equation
\begin{equation}\label{eq1}
i\frac{\partial \psi(x,y,z)}{\partial z} = \frac{1}{2} \left( \frac{\partial^2}{\partial x^2} + \frac{\partial^2}{\partial y^2} \right) \psi(x,y,z) ,
\end{equation}
where $\psi$ is the envelope of the beam structure.
The general solution of Eq. (\ref{eq1}) can be written as
\begin{equation}\label{eq2}
\psi(x,y,z) = {\mathcal{F}}^{-1} \left\{\exp\left( \frac{i}{2}(k_x^2+k_y^2) z\right){\mathcal{F}\{\psi(x,y,z=0)\}}\right\} ,
\end{equation}
where $k_x$ and $k_y$ are the spatial frequencies, $\mathcal F$ and ${\mathcal F}^{-1}$ are the direct and inverse Fourier transform operators.
Clearly, for a given beam structure, the beam structure at any propagation distance $z$ can be obtained numerically.

\begin{figure}[htbp]
\centering
\includegraphics[width=0.6\columnwidth]{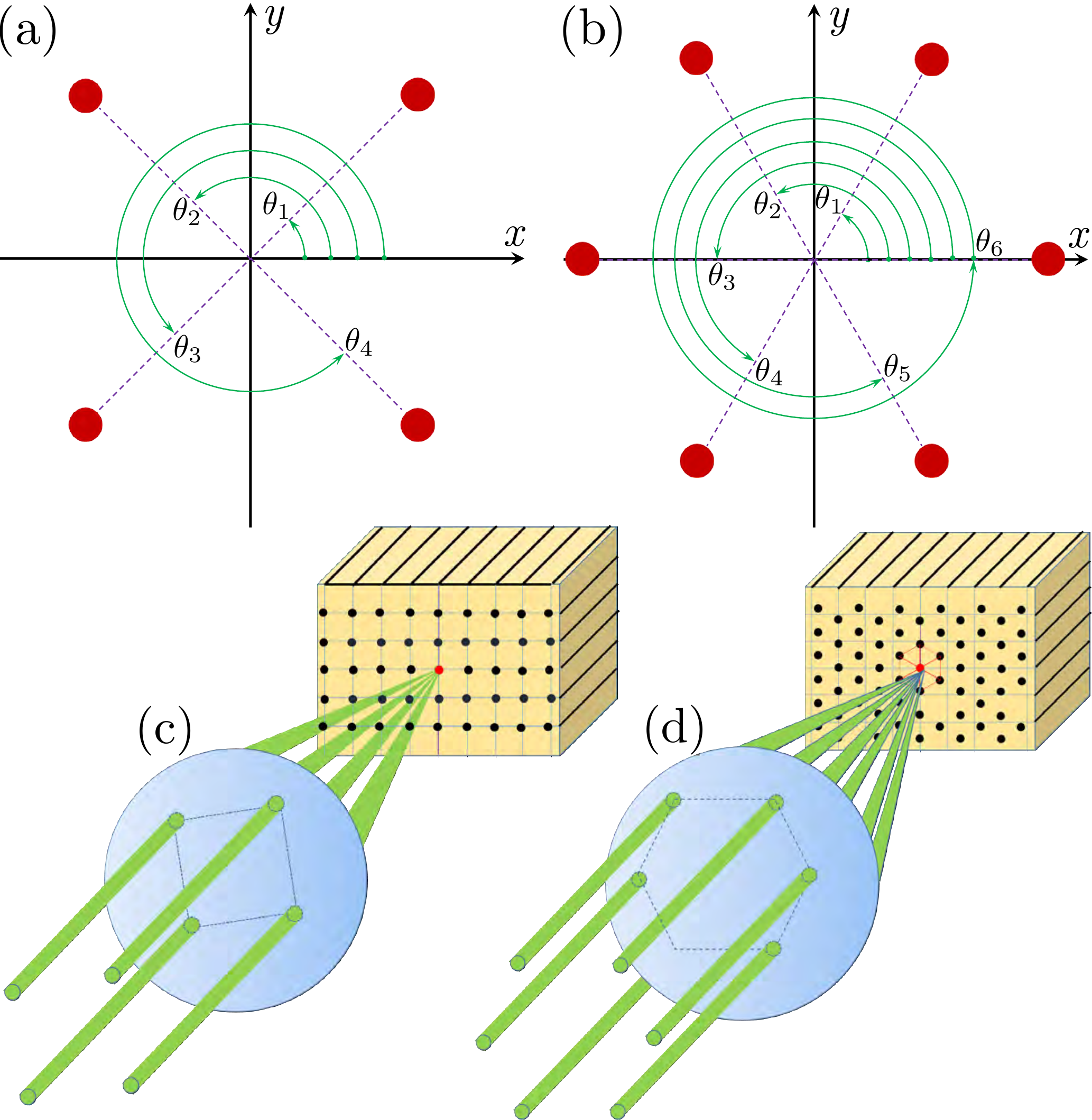}
\caption{(Color online) Illustration of the four-beam (a) and six-beam (b) interference in Cartesian coordinates.
Red dots represent the component beams.
(c,d) Three-dimensional schematics, corresponding to (a,b).}
\label{Fig1}
\end{figure}

The incident periodic beam structure can be obtained in many ways, and we opt for the multi-beam interference that can be written as
\begin{equation}
\psi(x,y)=\left| \sum_{m=1}^M \exp\left(i[x\cos(\theta_m)+y\sin(\theta_m)]+i\frac{ml}{M}\pi\right) \right|,
\end{equation}
where $M$ is the number of beams that are involved in interference, $\theta_m=2m\pi/M$ indicates the location of the beam,
and $l$ adjusts the initial phase shift.
Here, the absolute sign guarantees that only the beam envelope is modulated, so that the beam is not diffractionless any more.
In Fig. \ref{Fig1}, the four-beam interference geometry [Figs. \ref{Fig1}(a) and \ref{Fig1}(c)] and the six-beam interference geometry [Figs. \ref{Fig1}(b) and \ref{Fig1}(d)] are presented.
For the four-beam interference, the incident periodic structure is written as
\begin{align}\label{eq4}
\psi_{l=0} = &4 | \cos ({x}/{\sqrt{2}}) \cos ({y}/{\sqrt{2}})|,\\
\psi_{l=1} = &2 \sqrt{\sqrt{2} [1+\cos (X) \sin(Y)]+\cos(X)+\sin(Y)},\\
\psi_{l=2} = & \sqrt{1-\cos(X) \cos(Y)},\\
\psi_{l=3} = & 2 \sqrt{\sqrt{2} [1-\cos (X) \sin(Y)] -\cos(X)+\sin(Y)},\\
\psi_{l=4} = & 4 | \sin({x}/{\sqrt{2}}) \sin ({y}/{\sqrt{2}})|,
\end{align}
with $X=\sqrt{2}x$ and $Y=\sqrt{2}y$.
While for the six-beam interference, the corresponding incident periodic beam structure can be written as
\begin{align}\label{eq5}
\psi_{l=0} = &2 | 2 \cos ({x}/{2}) \cos ({\sqrt{3} y}/{2})+\cos (x)| , \\
\psi_{l=6} = &4 | [\cos ({x}/{2})-\cos ({\sqrt{3} y}/{2})] \sin ({x}/{2})|.
\end{align}
Here,  only the cases with $l=0$ and $l=6$ are presented, since the expressions for other cases are too complex.
Actually, as explained in the following text, the two cases presented ($l=0$ and 6) suffice to generate the super-honeycomb lattice.
Similar setups were employed previously \cite{boguslawski.pra.84.013832.2011,boguslawski.apl.98.061111.2011,terhalle.prl.101.013903.2008},
so we believe that our results are feasible for actual experiments.

\section{Results}
\subsection{Four-beam interference}

\begin{figure}[htbp]
\centering
\includegraphics[width=0.6\columnwidth]{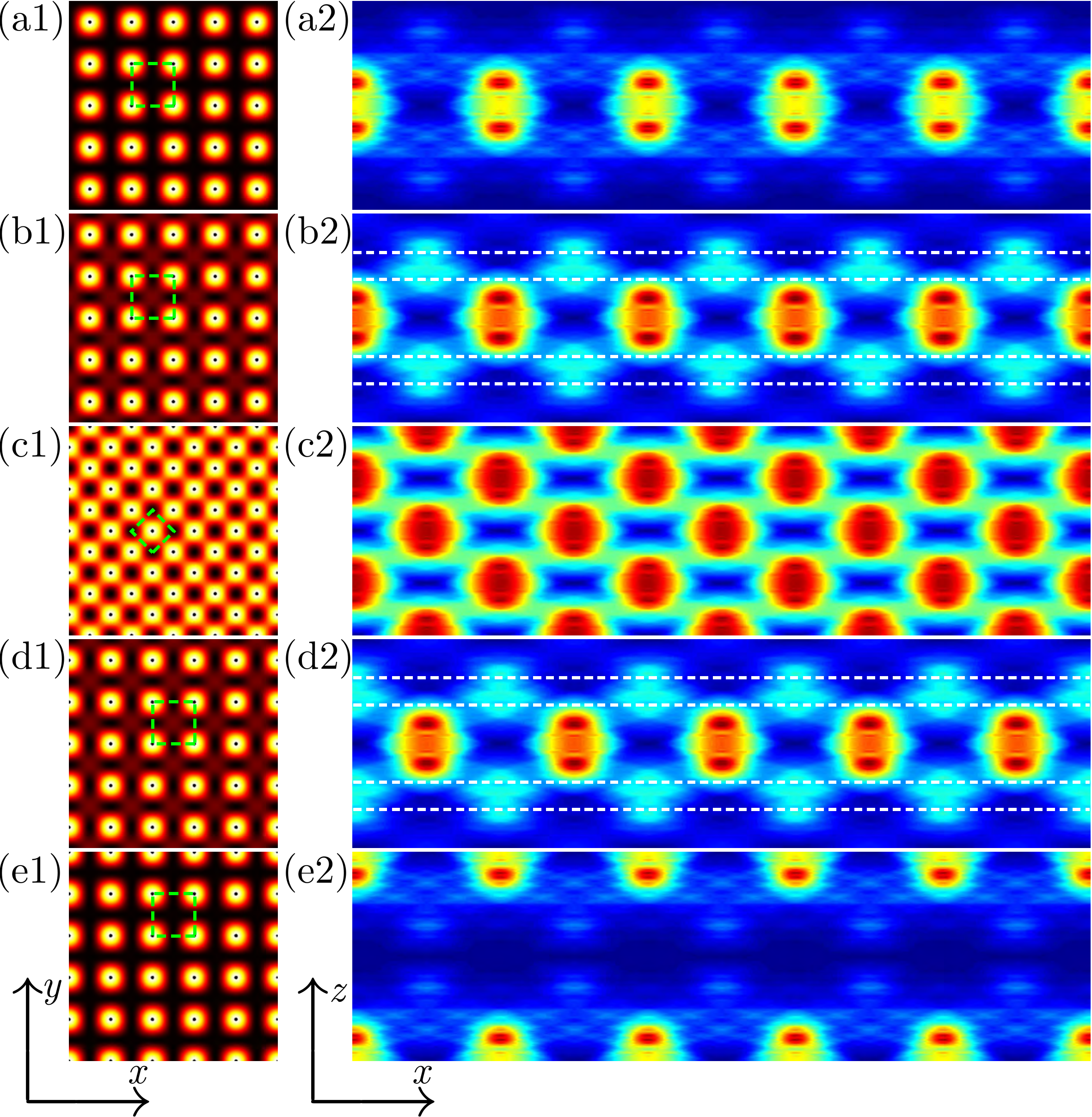}
\caption{(Color online) (a1)-(e1) Periodic beam structures (square lattices) due to four-beam interference with $l=0,\,1,\,2,\,3$ and 4, respectively. Dashed squares indicates unit cells.
Distributions are shown for $x\in[-2.5D_x,2.5D_x]$ and $y\in[-2.5D_y,2.5D_y]$.
(a2)-(e2) The corresponding Talbot images. Here, the variance of the beam is only recorded at a certain $y$ value $y=-2.5D_y$.
Distributions are shown for $x\in[-2.5D_x,2.5D_x]$ and $z\in[0,2\pi]$.
Dashed lines in (b2) and (d2) indicate the distances where the Lieb lattices can be observed.
{In (a1)-(e1), dark and bright regions represent low and high intensities.
In (a2)-(e2), blue and red regions represent low and high intensities.} }
\label{Fig2}
\end{figure}

First, we consider the four-beam interference case.
In Figs. \ref{Fig2}(a1)-\ref{Fig2}(e1), the incident periodic structures due to four-beam interference are exhibited, by changing the value of $l$ from 0 to 4 .
After some algebra, one finds that the transverse period of the beam structure is $D_x=D_y=\sqrt{2}\pi$, except for the case in Fig. \ref{Fig2}(c1),
in which $D_x=D_y=\pi$.
{According to Eqs. (\ref{eq4}) and (\ref{eq5}), which are described by the triangle functions, the periods
of the structures can be trivially obtained.}
The dashed squares in these panels indicate unit cells of the beam structure.
Thus, the total Talbot length for a two-dimensional structure is the least common multiple {(LCM)} of the two transverse Talbot lengths, i.e., $z_T={\rm LCM}(z_{Tx},z_{Ty})$,
which is $2\pi$ for the cases in Figs. \ref{Fig2}(a1)-\ref{Fig2}(e1) and $\pi$ for the case in Fig. \ref{Fig2}(c1).
In Figs. \ref{Fig2}(a2)-\ref{Fig2}(e2), the corresponding Talbot images of the periodic beam structures are shown within the propagation distance $0\le z\le 2\pi$.
For convenience, only $\psi(x,y=-2.5D_y)$ is recorded during propagation.
Clearly, one finds that the Talbot length in Fig. \ref{Fig2}(c2) is indeed half of that of the other cases.

Particularly, for the cases in Figs. \ref{Fig2}(b2) and \ref{Fig2}(d2), one finds that the incident beam structure changes into a Lieb lattice at
$z \approx 0.185z_T,\,0.315z_T,\,0.685z_T,$ and $0.815z_T$.
Such places are marked by white dashed lines in Figs. \ref{Fig2}(b2) and \ref{Fig2}(d2).

\begin{figure}[htbp]
\centering
\includegraphics[width=0.6\columnwidth]{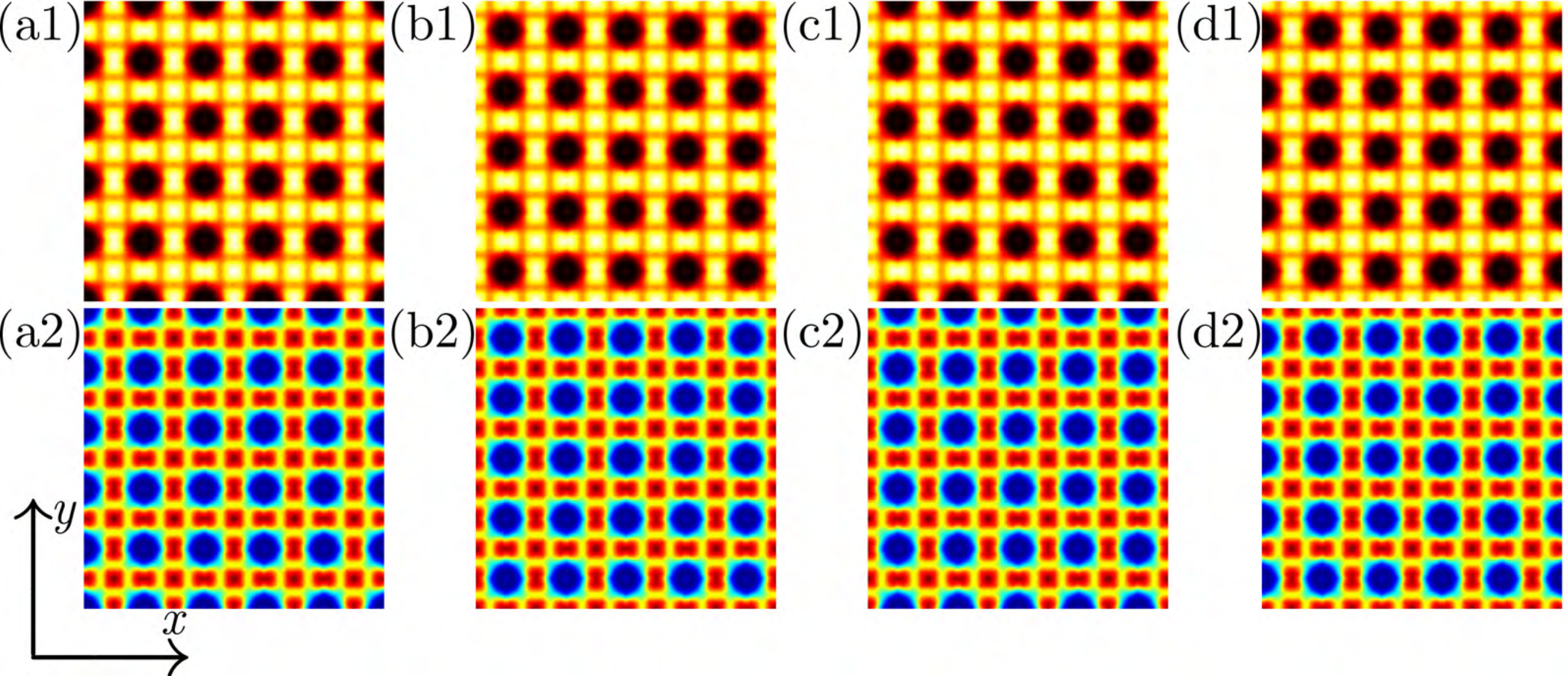}
\caption{(Color online) (a1)-(d1) Lieb lattices obtained numerically.
(a2)-(d2) Lieb lattices obtained according to Eq. (\ref{eq2}).
(a,b) and (c,d) correspond to $z=(0.185z_T,0.315z_T)$ {[dashed lines in Figs. \ref{Fig2}(b2) and \ref{Fig2}(d2)]}, respectively. 
}
\label{Fig3}
\end{figure}

In Fig. \ref{Fig3}, the found Lieb lattices are displayed,
in which the results in the first row [Figs. \ref{Fig3}(a1)-\ref{Fig3}(d1)] are obtained numerically,
while those in the second row [Figs. \ref{Fig3}(a2)-\ref{Fig3}(d2)] are obtained analytically, according to Eq. (\ref{eq2}).
One finds that the numerical and analytical results completely agree with each other.

\subsection{Six-beam interference}

\begin{figure}[htbp]
\centering
\includegraphics[width=0.6\columnwidth]{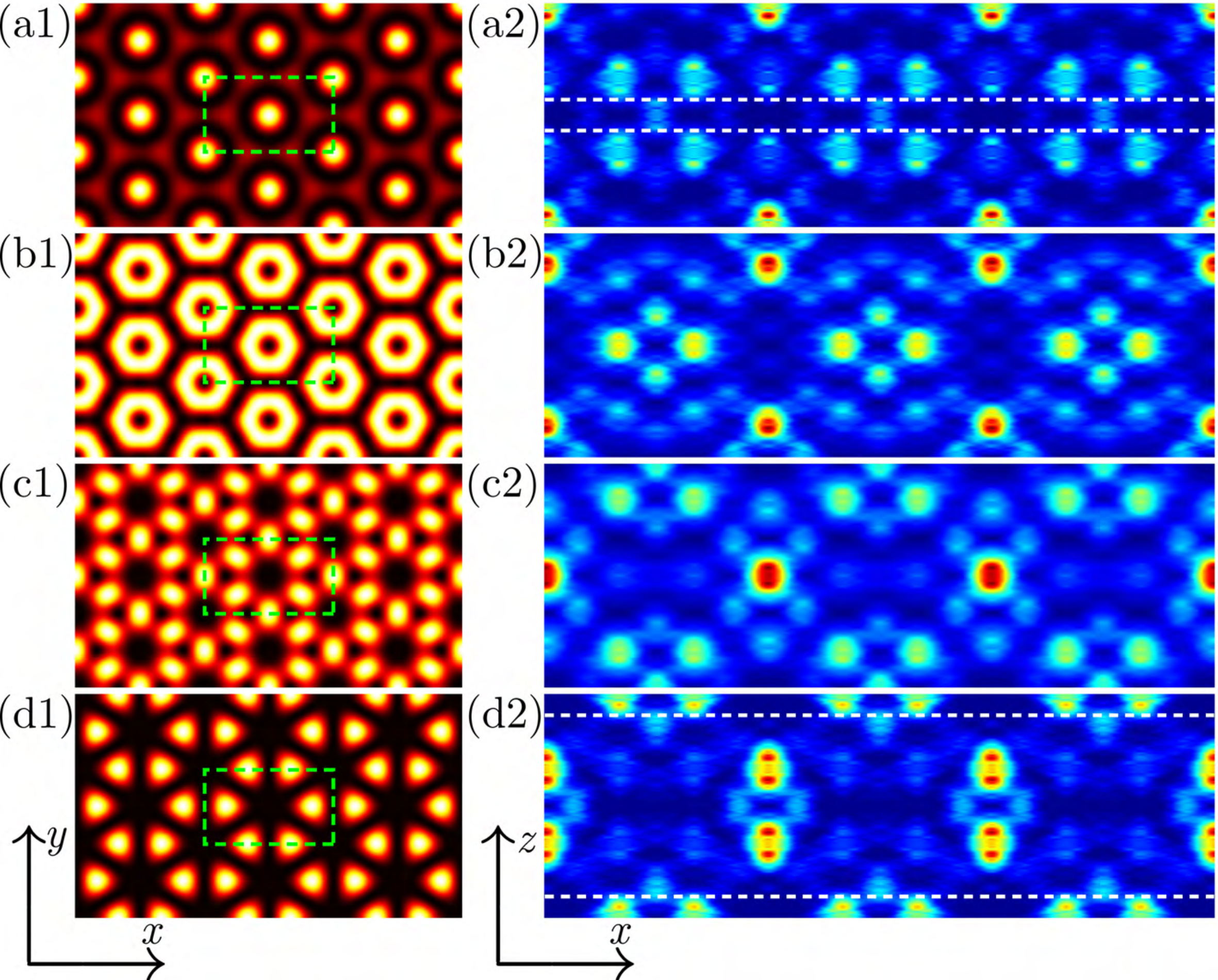}
\caption{(Color online) Setup is as in Fig. \ref{Fig2}, but appropriate for the six-beam interference and for the observation of super-honeycomb lattices.
(a1)-(d1) $l=0,\,2,\,4$ and 6, respectively.
Distributions in (a1)-(d1) are shown for $x\in[-1.5D_x,1.5D_x]$ and $y\in[-1.5D_y,1.5D_y]$,
in (a2)-(d2) are shown for $x\in[-1.5D_x,1.5D_x]$ and $z\in[0,z_T]$.
Dashed rectangles in (a1)-(d1) represent the recurring cells.}
\label{Fig4}
\end{figure}

Second, we treat the six-beam interference case, as shown in Fig. \ref{Fig1}(b).
Similar to Fig. \ref{Fig2}, the incident periodic beam structure adjusted by $l$ and the corresponding Talbot images are exhibited in Fig. \ref{Fig4}.
In Figs. \ref{Fig4}(a1)-\ref{Fig4}(d1), the dashed rectangles indicate the recurring cells (not the unit cells).
For this case, $D_x=2\pi$ and $D_y=2\pi/\sqrt{3}$, so the Talbot length is $z_T=4\pi$.
In Fig. \ref{Fig4}(a1) with $l=0$, the beam structure shows a mixture of hexagonal and honeycomb lattices.
In Fig. \ref{Fig4}(b1) with $l=2$, the bright rings form the hexagonal lattice.
If one launched this pattern into a self-defocusing medium, a mixture of super-honeycomb lattice and hexagonal lattices can be obtained \cite{lan.prb.85.155451.2012}.
If $l=4$ and 6, the interference patterns are the kagome [Fig. \ref{Fig4}(c1)] and honeycomb [Fig. \ref{Fig4}(d1)] lattices \cite{boguslawski.apl.98.061111.2011,boguslawski.pra.84.013832.2011,zong.oe.24.8877.2016}, respectively.
To display the Talbot effect clearly, we only recorded the beam at a certain $y$ value, $y=-1.5D_y$, during propagation, as in Figs. \ref{Fig4}(a2)-\ref{Fig4}(d2).

Numerical simulations demonstrate that the super-honeycomb lattices may form only if $l=0$ or $l=6$.
For the former case, one may observe the super-honeycomb lattice at $z\approx0.431z_T$ and $0.569z_T$,
while for the latter case, the typical distances are $z\approx0.095z_T$ and $0.905z_T$.
These typical distances are marked by the dashed lines in Figs. \ref{Fig4}(a2) and \ref{Fig4}(d2).

\begin{figure}[htbp]
\centering
\includegraphics[width=0.6\columnwidth]{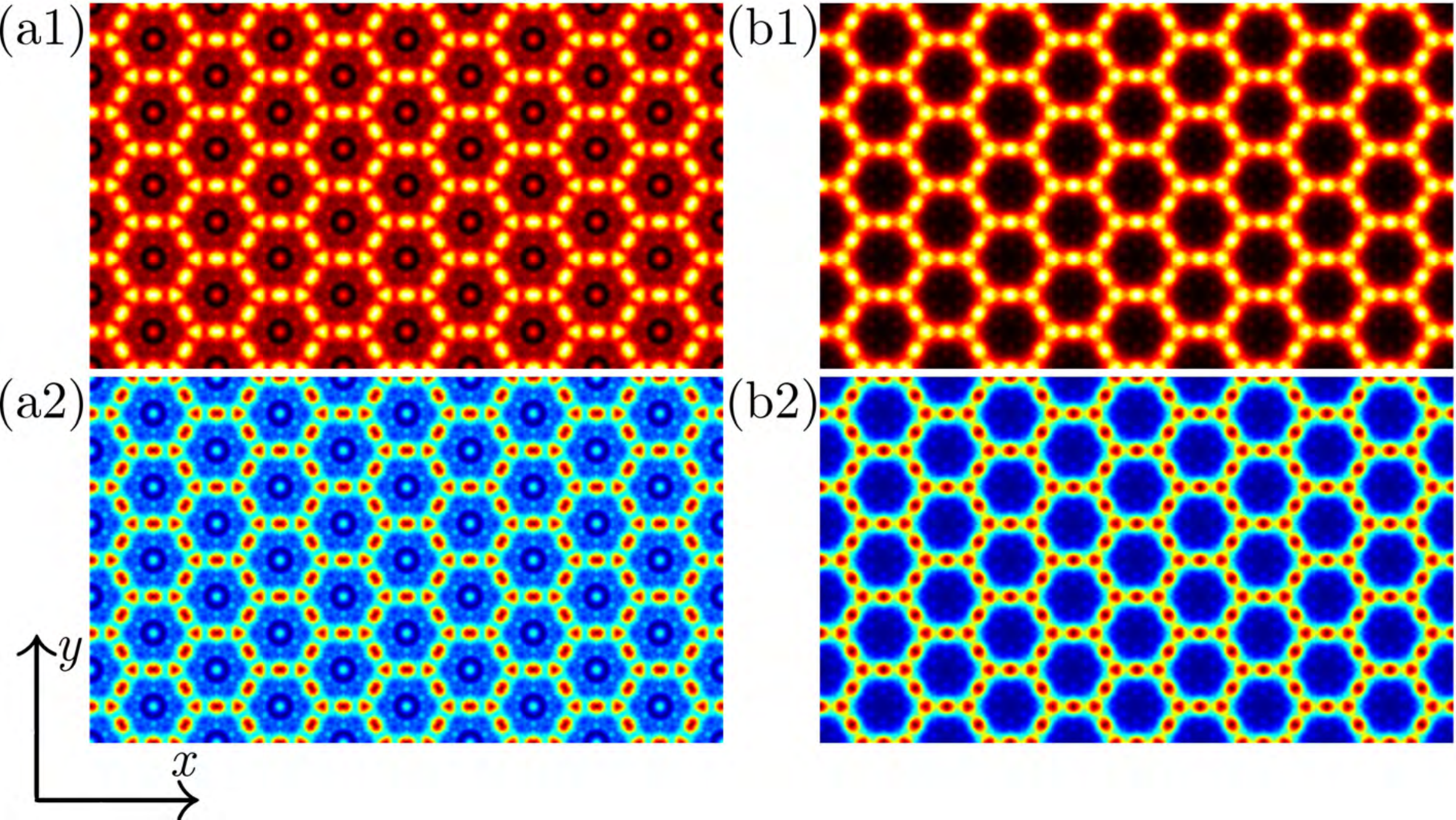}
\caption{(Color online) (a1) and (b1) Super-honeycomb lattices obtained numerically.
(a2) and (b2) Super-honeycomb lattices obtained analytically, according to Eq. (\ref{eq2}).
(a) and (b) correspond to $z=(0.431z_T,0.095z_T)$ {[dashed lines in Figs. \ref{Fig4}(b2) and \ref{Fig4}(d2)]}.
Distributions are shown for $x\in[-2.5D_x,2.5D_x]$ and $y\in[-2.5D_y,2.5D_y]$.}
\label{Fig5}
\end{figure}

The obtained super-honeycomb lattices are shown in Fig. \ref{Fig5}.
The numerical results in the first row and analytical results in the second row agree with each other very well.
For the case with $l=0$ [Figs. \ref{Fig5}(a1) and \ref{Fig5}(a2)], the pattern is a mixture of super-honeycomb and hexagonal lattices,
but the strength of the hexagonal lattice is lower than that of the super-honeycomb lattice.
Therefore, one can improve the resolution and contrast by applying a high-pass filter.
Nevertheless, the case is still much better than the result obtained through launching the pattern in Fig. \ref{Fig4}(b1) into a self-defocusing medium \cite{lan.prb.85.155451.2012},
because high-pass filtering is not valid for this situation.
One can only use an additional beam structure to balance the hexagonal lattice, which is inconvenient.
For the case with $l=6$, the obtained super-honeycomb lattice is shown in Figs. \ref{Fig5}(b1) and \ref{Fig5}(b2).
In comparison with the results in Figs. \ref{Fig5}(a1) and \ref{Fig5}(a2), the quality of the super-honeycomb lattice in Figs. \ref{Fig5}(b1) and \ref{Fig5}(b2) is significantly improved.
Undoubtedly, the method to generate the super-honeycomb lattice is simple and effective.

\subsection{Lattices with broken inversion symmetry}

\begin{figure}[htbp]
\centering
\includegraphics[width=0.6\columnwidth]{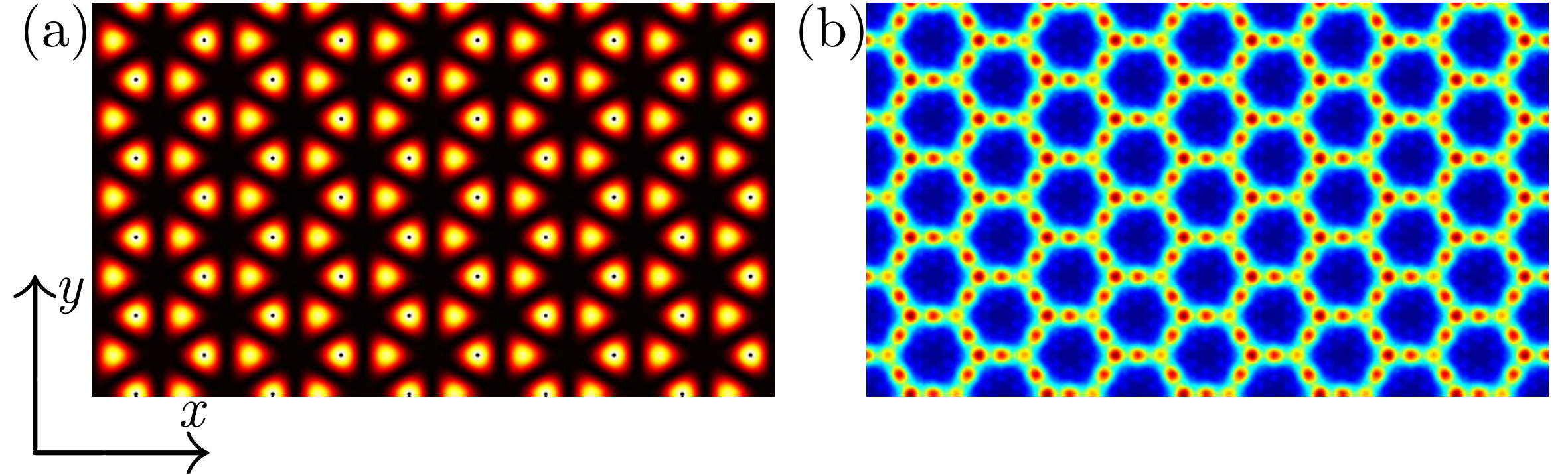}
\caption{(Color online) (a) Honeycomb lattice resulting from the six-beam interference with $l=5.8$.
(b) Super-honeycomb lattice obtained based on the inversion symmetry broken honeycomb lattice at $z=0.095z_T$.
Distributions are shown for $x\in[-2.5D_x,2.5D_x]$ and $y\in[-2.5D_y,2.5D_y]$.}
\label{Fig6}
\end{figure}

In the examples above, the value of $l$ is an integer. What happens if $l$ is non-integer?
To find the answer, we consider six-beam interference with $l=5.8$, as an example.
The induced periodic structure is shown in Fig. \ref{Fig6}(a); one finds that the inversion symmetry of the honeycomb lattice is broken.
The broken inversion symmetry will open Dirac cones in the honeycomb lattice in the momentum space and lead to two kinds of valleys with opposite Berry curvatures.
Recently, lattices with broken inversion symmetry were involved in investigating the valley Hall effect and topological insulators with invariant time-reversal symmetry
\cite{dong.nm.16.298.2016,wu.nc.8.1304.2017,lu.np.13.369.2017,noh.prl.120.063902.2018}.
Based on such incident beam structure, the formed super-honeycomb lattice is shown in Fig. \ref{Fig6}(b).
Apparently, the inversion symmetry of the super-honeycomb lattice is broken.

\section{Discussion}
We would like to emphasize that the results obtained in this paper should not be simply viewed as a collection of beautiful patterns
that can be obtained by superposing different beams with different phases.
The Lieb as well as the super-honeycomb structure are not obtained arbitrarily, but as a result of a well-designed precise procedure based on the fractional Talbot effect.
The value of this research is in providing a simple and more direct method to construct Lieb and super-honeycomb lattices, for example
by choosing proper parameters of the superposed beams in a 4f-system.
Due to the complexity of integrals in Eq. (\ref{eq2}),
the fractional Talbot length for observing such lattices cannot be analytically obtained.
However, numerical simulations can predict such lengths quite well, without affecting the consequences of the method.

The method utilizes effectively the self-imaging effect, and is based only on the multi-beam interference.
Therefore, it is not difficult to implement such patterns and induce the corresponding lattices in photorefractive and other dielectric media \cite{terhalle.prl.101.013903.2008}.
It should be noted that the results reported in this paper can be experimentally verified by utilizing the interference among the so-called pseudo-nondiffracting beams, as previously used in \cite{boguslawski.apl.98.061111.2011,boguslawski.pra.84.013832.2011}.
Last but not least, the method may also be used for producing lattices in ultracold gases \cite{tarruell.nature.483.302.2012,jotzu.nature.515.237.2014},
due to formal equivalence between the Schr{\"o}dinger equation and the paraxial wave equation;
the propagating distance $z$ in latter equation plays the role of evolution time $t$ in the former equation.

\section{Conclusion}

In summary, this paper has reported a new method to produce Lieb and super-honeycomb lattices with high quality,
based on the fractional Talbot effect.
By properly choosing the initial phase shifts of the interfering beams,
the periodic beam structure has been shown to form the Lieb or super-honeycomb lattices during propagation, at certain fractional Talbot lengths.
Last but not least, the inversion symmetry of the incident as well as the induced periodic beam structures can be broken by adjusting the initial phase shifts of the interfering beams.
This method is relatively direct and can be {\it in situ} manipulated.
On one hand, this research broadens the applications of  the Talbot effect and deepens its understanding,
and on the other, it may provide a useful reference for research in topological photonics.

\section*{Funding}
Natural Science Foundation of Shaanxi Province (2017JZ019);
Natural Science Foundation of Guangdong Province (2018A0303130057);
Qatar National Research Fund (NPRP 8-028-1-001).

\bibliography{my_refs_library}

\end{document}